\RequirePackage{rotating}
\documentclass[usenatbib]{mnras}
\usepackage{epsfig}
\usepackage{amsmath}
\usepackage{graphicx}
\usepackage{array}
\usepackage{textcomp}
\usepackage{amssymb}
\usepackage{rotating}
\usepackage{pdflscape}
\usepackage{caption}
\usepackage{subcaption}
\usepackage{longtable}
\usepackage{xtab}
\def\mnras {{\sc MNRAS}}
\def\apj {ApJ}

\def\apjl {ApJL}
\def\aj {AJ}

\def\aap {A\&A}

\title[Stellar Populations of Bars]
      {SDSS-IV MaNGA: Stellar Population Gradients Within Barred Galaxies.} 
\author[A.\ Fraser-McKelvie et al.]
       {Amelia Fraser-McKelvie$^{1}$\thanks{Amelia.Fraser-McKelvie@nottingham.ac.uk}, Michael Merrifield$^{1}$, Alfonso Arag\'{o}n-Salamanca$^{1}$, 
      \and Thomas Peterken$^{1}$, Karen Masters$^{2,3}$, Coleman Krawczyk$^{3}$,  Brett Andrews$^{4}$, \and Johan H.~Knapen$^{5,6,7}$, Sandor Kruk$^{8}$,
     Adam Schaefer$^{9}$, Rebecca Smethurst$^{8}$, \and Rog\'{e}rio Riffel$^{10,11}$, Joel Brownstein$^{12}$, Niv Drory$^{13}$       
        \vspace*{1mm}\\
        $^{1}$ School of Physics \& Astronomy, University of Nottingham, University Park, Nottingham, NG7 2RD, U.K. \\
        $^{2}$ Department of Physics and Astronomy, Haverford College, 370 Lancaster Ave, Haverford, PA 19041, U.S.A. \\
        $^{3}$ Institute of Cosmology \& Gravitation, University of Portsmouth, Dennis Sciama Building, Portsmouth, PO1 3FX, U.K. \\
        $^{4}$ PITT PACC, Department of Physics and Astronomy, University of Pittsburgh, Pittsburgh, PA 15260, U.S.A. \\
        $^{5}$ Instituto de Astrof\'isica de Canarias, 38205 La Laguna, Tenerife, Spain\\
        $^{6}$ Departamento de Astrof\'isica, Universidad de La Laguna, E-38205, La Laguna, Tenerife, Spain\\
        $^{7}$ Astrophysics Research Institute, Liverpool John Moores University, IC2, Liverpool Science Park, 146 Brownlow Hill, Liverpool, L3 5RF, U.K.\\
        $^{8}$ Department of Astrophysics, University of Oxford, Denys Wilkinson Building, Keble Road, Oxford, OX1 3RH, U.K. \\
        $^{9}$ Department of Astronomy, University of Wisconsin-Madison, 475N. Charter St., Madison WI 53703, U.S.A. \\
        $^{10}$ Departamento de Astronomia, Universidade Federal do Rio Grande do Sul - Av. Bento Gon\c calves 9500, Porto Alegre, RS, Brazil.\\
        $^{11}$ Laborat\'orio Interinstitucional de e-Astronomia, Rua General Jos\'e Cristino, 77 Vasco da Gama, Rio de Janeiro, Brazil, 20921-400\\
        $^{12}$ Department of Physics and Astronomy, University of Utah, 115 S. 1400 E., Salt Lake City, UT 84112, U.S.A. \\
        $^{13}$ McDonald Observatory, The University of Texas at Austin, 1 University Station, Austin, TX 78712, U.S.A. \\
	}
\begin{document}
\maketitle
\begin{abstract}

Bars in galaxies are thought to stimulate both inflow of material and radial mixing along them. Observational evidence for this mixing has been inconclusive so far however, limiting the evaluation of the impact of bars on galaxy evolution. We now use results from the MaNGA integral field spectroscopic survey to characterise radial stellar age and metallicity gradients along the bar and outside the bar in 128 strongly barred galaxies. We find that age and metallicity gradients are flatter in the barred regions of almost all barred galaxies when compared to corresponding disk regions at the same radii. Our results re-emphasize the key fact that by azimuthally averaging integral field spectroscopic data one loses important information from non-axisymmetric galaxy components such as bars and spiral arms. We interpret our results as observational evidence that bars are radially mixing material in galaxies of all stellar masses, and for all bar morphologies and evolutionary stages.

\end{abstract}
\section{Introduction}

Galactic bars are long-lived phenomena \citep{Gadotti15} that occur in a large fraction of local Universe disk galaxies \citep[e.g.][]{Knapen00, Masters11}. 
Simulations show gas and angular momentum may be funnelled along bars to the central regions of a galaxy \citep[e.g.][]{Simkin80,Weinberg85,Knapen95,Minchev10,Brunetti11, Spinoso17}, and bar-driven secular evolution is a likely candidate for the cessation of star formation in galaxies at late times \citep[e.g.][]{Masters12, Kruk18}.

The orbits of stars within bars have been well studied \citep[e.g.][]{Combes81, Sellwood81,Athanassoula92}. The classical picture of bar formation and evolution requires the majority of stars in bar regions to be trapped around periodic elongated orbits in the direction of the bar major axis, known as the $x_{1}$ class of orbits \citep{Contopoulos80}.
If we assume that bar stellar orbits are indeed elongated with respect to stellar orbits within the disk of the galaxy, the bar may be treated as a confined structure within a galaxy. It follows that we would expect a greater radial mixing of stellar populations within bars if they formed from the same population as the disk. This would manifest itself observationally as weaker stellar age and metallicity gradients within the bar regions compared to non-bar regions of the galaxy at the same radii. We will refer to this mixing of stellar populations within a bar as `radial mixing', but note this is a separate phenomenon to the radial mixing observed in disks of barred galaxies outside of co-rotation, as described in previous literature \citep[e.g.][]{Friedli94, DiMatteo13}.

Observations of the direct effect of a bar on its host galaxy have been diverse in their approach, and have produced contrasting results. Single-fibre studies of the stellar populations of central (bulge-dominated) regions of barred galaxies have been shown to be no different to those of non-barred galaxies \citep{Cacho14}. In addition, azimuthally-averaged gradients over inner regions are also comparable to non-barred galaxies \citep{Cheung15}.
Outside of co-rotation, simulations predict a flattening of stellar population gradients \citep{Friedli94, Minchev10} thought to be due to the resonant coupling between bars and spiral arms. This was not reproduced in observational results, however \citep{SB14}. It is possible that in averaging across an entire galaxy, any subtle differences in the stellar populations as a result of the presence of a bar is lost.

Recent long-slit and integral-field spectroscopic results have begun to extricate the bar component and attempt to treat it as a separate entity within a galaxy.
Long slit works such as those by \citet{Perez07} and \citet{Perez09} placed slits along the bar major axes of a small sample of galaxies, reporting differences in stellar population gradient trends. For a sample of 20 galaxies, the authors report bars can possess positive, negative, or no metallicity gradient, and these gradients are correlated with galaxy velocity dispersion and mean stellar age. \citet{SB11} reanalysed the observations of \citet{Perez09} to include the disk regions of two barred galaxies, and found the bars of these galaxies contained flatter age and metallicity gradients. In a novel approach, \citet{Williams12} measure stellar population gradients of the central regions of 22 edge-on disk galaxies with boxy-peanut bulges (indicating the presence of a bar), and find flattened stellar population gradients when compared to non-barred galaxies.
It seems that spatial information is crucial in determining stellar population trends across galaxies that contain non-axisymmetric structures.
Indeed, when spatially-resolved integral field spectroscopy is considered, \citet{Seidel16} in a pilot study confirm a telltale flattening of stellar metallicity gradients on average along bars compared to disk regions of the same galaxy in 16 galaxies. The above results are consistent with bar mixing, though all have been derived from studies of small samples of galaxies. Whether this trend holds for all barred galaxies, and through all stages of bar evolution, is not known. 

It is clear that in order to facilitate a detailed analysis of stellar population gradient trends, a large sample of barred galaxies, spatially-resolved spectroscopy, and bar positional information within the galaxy are required. All three of these conditions are met by using the Mapping Nearby Galaxies at APO (MaNGA) galaxy survey in conjunction with the new citizen science project, Galaxy Zoo:3D, which aims to separate light from structural galaxy components within MaNGA data cubes. With this in mind, in this letter we examine the stellar populations within the bar and disk regions of a large sample of local-Universe barred galaxies using integral field spectroscopic (IFS) data from the MaNGA galaxy survey. In Section~\ref{sample} we describe the MaNGA survey and the barred galaxy sample used, in Section~\ref{method} we detail how the stellar population indicators are measured, and in Section~\ref{results} we present the results.


\section{Sample}
\label{sample}
\subsection{The MaNGA Galaxy Survey}
The MaNGA Galaxy Survey is an integral field spectroscopic survey that aims to observe 10,000 galaxies by 2020 \citep{Bundy15, Drory15}. It is an SDSS-IV project \citep{Blanton17}, employing the 2.5m telescope at Apache Point Observatory \citep{Gunn06} and BOSS spectrographs \citep{Smee13}. MaNGA Product Launch 7 (MPL-7) contains 4620 unique galaxy observations, observed and reduced by the MaNGA Data Reduction Pipeline \citep{Law16}, with derived properties produced by the MaNGA Data Analysis Pipeline \citep[DAP;][]{Westfall19}, provided as a single data cube per galaxy \citep{Yan16}. MaNGA's target galaxies were chosen to include a wide range of galaxy masses and colours, over the redshift range $0.01<z<0.15$, and the Primary+ sample \citep[][]{Yan16b, Wake17} contains spatial coverage out to $\sim$1.5 $R_{\textrm{e}}$ for $\sim$66\% of all observed galaxies.

\subsection{Barred Galaxy Sample}
We select barred galaxies within the MaNGA MPL-7 sample using Galaxy Zoo 2 \citep{Hart16}, which is a citizen science project that provides morphological classifications for all MaNGA galaxies. We select galaxies with a weighted bar vote fraction of $>0.7$, which is the fraction of respondents that classified a particular galaxy as possessing a bar, weighted by participant agreement level with other users. From the 4620 galaxies in MPL-7, 488 are thus classified as highly likely to contain bars. We note that this is $\sim10\%$ of the original sample, and that while we are confident we have selected barred galaxies, these will be the strongest barred galaxies in the MaNGA sample. 

\begin{figure*}
\centering
\begin{subfigure}{0.99\textwidth}
\includegraphics[width=\textwidth]{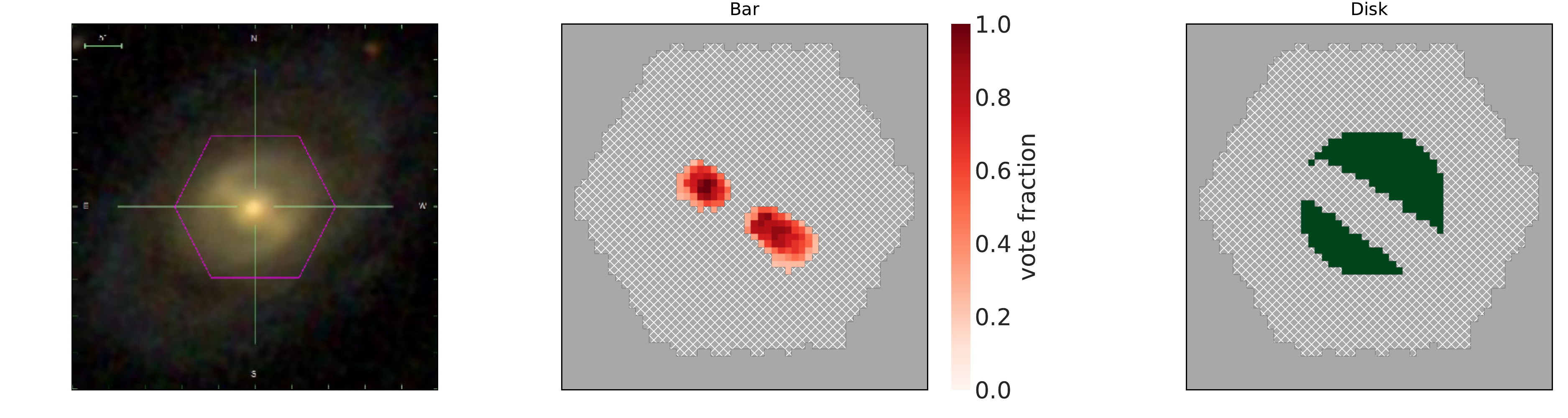} 
\end{subfigure}
\caption{The barred galaxy 8451-6101 (left, with MaNGA field of view as the pink hexagon), and its Galaxy Zoo:3D bar (red, centre) and disk (green, right) masks. The bar mask is scaled by the total vote fraction of respondents. Only spaxels with a vote fraction greater than 0.2 are included in the bar region for analysis. We exclude the central region of the galaxy in our gradient determination in order to compare bar and disk gradients at the same radii. The disk is defined as any spaxel within the IFU field of view that is not included in the bar region. }
\label{GZ3D_masks}
\end{figure*}

Apart from determining whether a galaxy possesses a bar, for spatially-resolved stellar population analysis of a large sample of galaxies, we also require an automated method to determine the region in the IFU datacube where the bar lies. For this, we use Galaxy Zoo:3D (GZ:3D; Masters et al, in prep.), a new citizen science project that asks participants to trace regions on a galaxy image that correspond to various components seen, including bars, spiral arms, and bulges. 
The regions drawn are translated into masks, weighted by the number of users that determined each spatial pixel (spaxel) to be located within the region of interest. Figure~\ref{GZ3D_masks} shows an example of the GZ:3D masks for the MaNGA galaxy 8451-6101.

The initial input sample into GZ:3D were galaxies that were likely to contain spiral arms according to Galaxy Zoo 2 from MPL-5 (2836 galaxies). While not all of these galaxies actually contain spiral arms, the large majority do, and hence our sample is biased towards spiral galaxies, with less than 15\% S0s. 
To date, GZ:3D has only been run on a portion of the MaNGA sample, and of the 488 barred galaxies in MPL-7, 128 also possess GZ:3D bar region masks. This final sample of 128 barred galaxies spans a mass range of $5.3\times10^{8}<\textrm{M}/ \textrm{M}_{\odot}<1.4 \times 10^{11}$. There is a slight bias in the sample used in this analysis towards lower-mass, optically-bluer galaxies than the overall barred galaxy population in MaNGA (a result of the low S0 fraction that possess GZ:3D masks), but differences in median colour and mass distributions are less than 1$\sigma$.

\section{Measuring Stellar Populations}
\label{method}
In order to compare the physical properties both within and outside the bar for a given radius, we extract the stellar populations of both the bar and disk regions of barred MaNGA galaxies using full spectral fitting. As a check, we also derive population properties from index measurements. We define the bar regions as any spaxel within the GZ:3D bar masks in which at least 80\% of respondents have determined this to be a bar region. We make this conservative cut to account for any respondents who drew spurious regions on the GZ:3D images. We define a corresponding disk region as all spaxels within the same radius as the original bar masks, but not within the bar mask. We note that spiral arms generally exist outside of a bar radius, and in the cases where they do contaminate our measured disk region, we ignore any secondary effects they may contribute for this analysis. It is important to compare regions within a galaxy at similar radii, and given our definition of the bar region, it is clear there are no associated disk regions in the central radii of a galaxy. For this reason, we exclude the central regions of the galaxy from this analysis, and choose only to look at regions of the galaxy that have both bar and disk spaxel mask regions. We note that stellar population gradients measured may differ from literature values as we are not including the central regions of the galaxy, but this ensures bar and disk stellar population parameters will be internally consistent with one another for a given galaxy.

We perform full spectral fitting using Starlight \citep{CidFernandes05}, and a subset of the E-MILES set of synthetic SSP templates \citep{Vazdekis16} on every spaxel in the MaNGA barred galaxy sample. We do not require spatial binning as we are rarely using spaxels in outer (lower signal-to-noise) regions of the IFUs, and need to avoid blending any signal into bins that incorporate both bar and disk regions. By assuming a Kroupa revised IMF \citep{Kroupa01}, BaSTI isochrones \citep{Pietrinferni04}, and a Milky Way $\alpha$/Fe, the best-fit spectrum is obtained along with the weighted combination of templates used, from which we derive average ages and metallicities for each spaxel.

For the index-derived stellar population estimates, we employ H$\beta$ as a stellar age indicator, as it is sensitive to the presence of young stars, and [MgFe]$^{\prime}$, defined by:
\begin{equation*}
[\textrm{MgFe}]^{\prime} = \sqrt{(\textrm{Mg}b (0.72 \times \textrm{Fe5270} + 0.28 \times \textrm{Fe5335})},
\end{equation*}
as a stellar metallicity indicator. Both these indices are relatively insensitive to changes in $\alpha$/Fe ratio \citep{Gonzalez93}. 
Using \textsc{ppxf} \citep{Cappellari17}, the MaNGA DAP fits a combination of stellar spectra to the MaNGA spectrum, then subtracts emission lines, measures absorption lines, and corrects for instrumental resolution and Doppler broadening effects \citep{Westfall19}. The resultant maps of absorption line indices are used in this work. We infer age and metallicity estimates from index measurements by interpolating over a grid of MILES SSP models of \citet{Vazdekis10}, scaled to the flux-weighted average velocity dispersion of the galaxy. 

\section{Results \& Discussion}
\label{results}
To measure the stellar population gradients, we first average all spaxel age and metallicity values in azimuthal rings of 0.5$^{\prime\prime}$ based on the light-weighted elliptical radius of each spaxel from the galaxy centre. This measurement takes in to account inclination effects so that regions at the same physical radius from the centre of a galaxy are compared to each other.
A linear least-squares fit was then performed to the Starlight- and index-derived age and metallicity values as a function of their (linear) distance to the galaxy centre to obtain the age and metallicity gradients.

\begin{figure*}
\centering
\begin{subfigure}{0.45\textwidth}
\includegraphics[width=\textwidth]{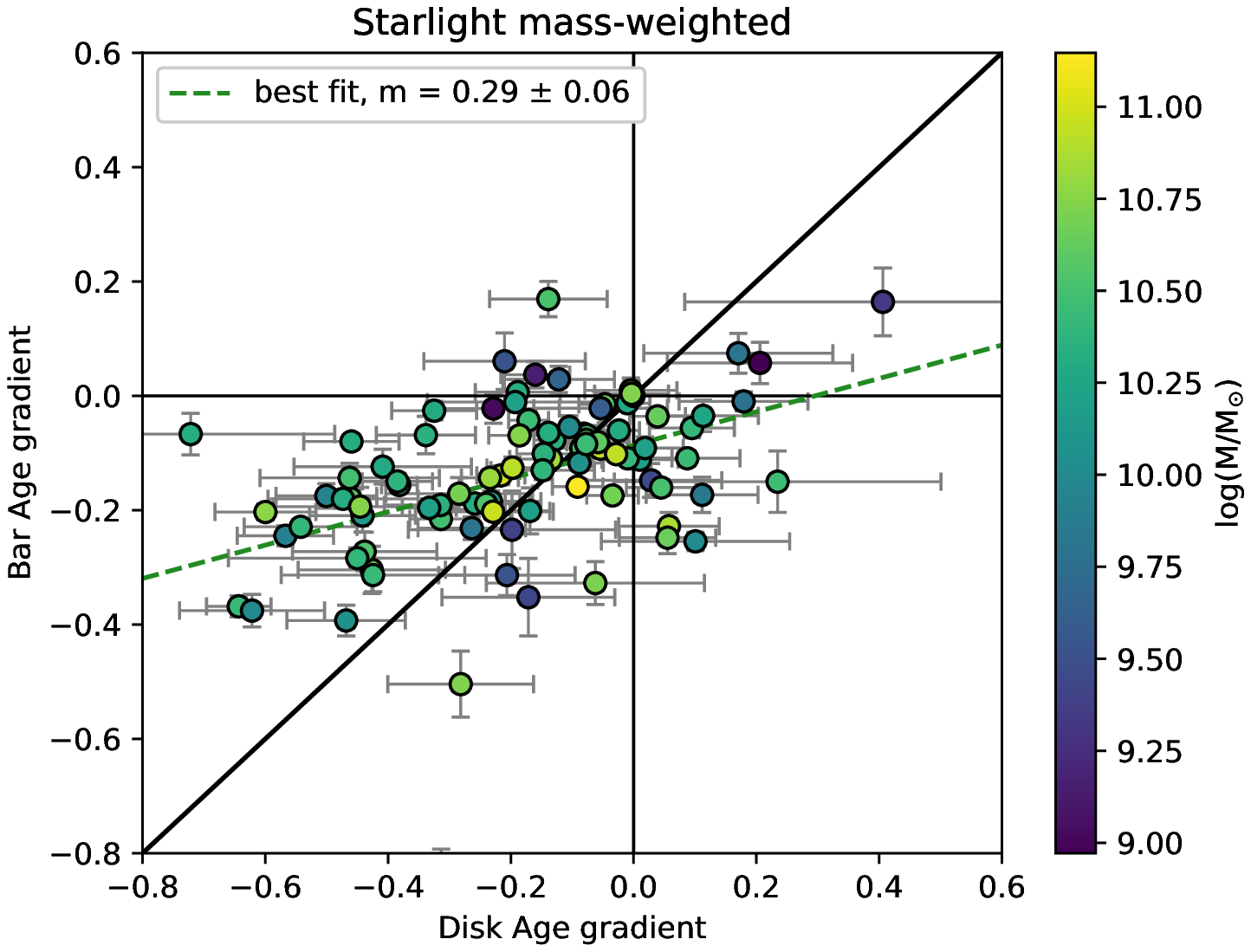} 
\end{subfigure}
\begin{subfigure}{0.45\textwidth}
\includegraphics[width=\textwidth]{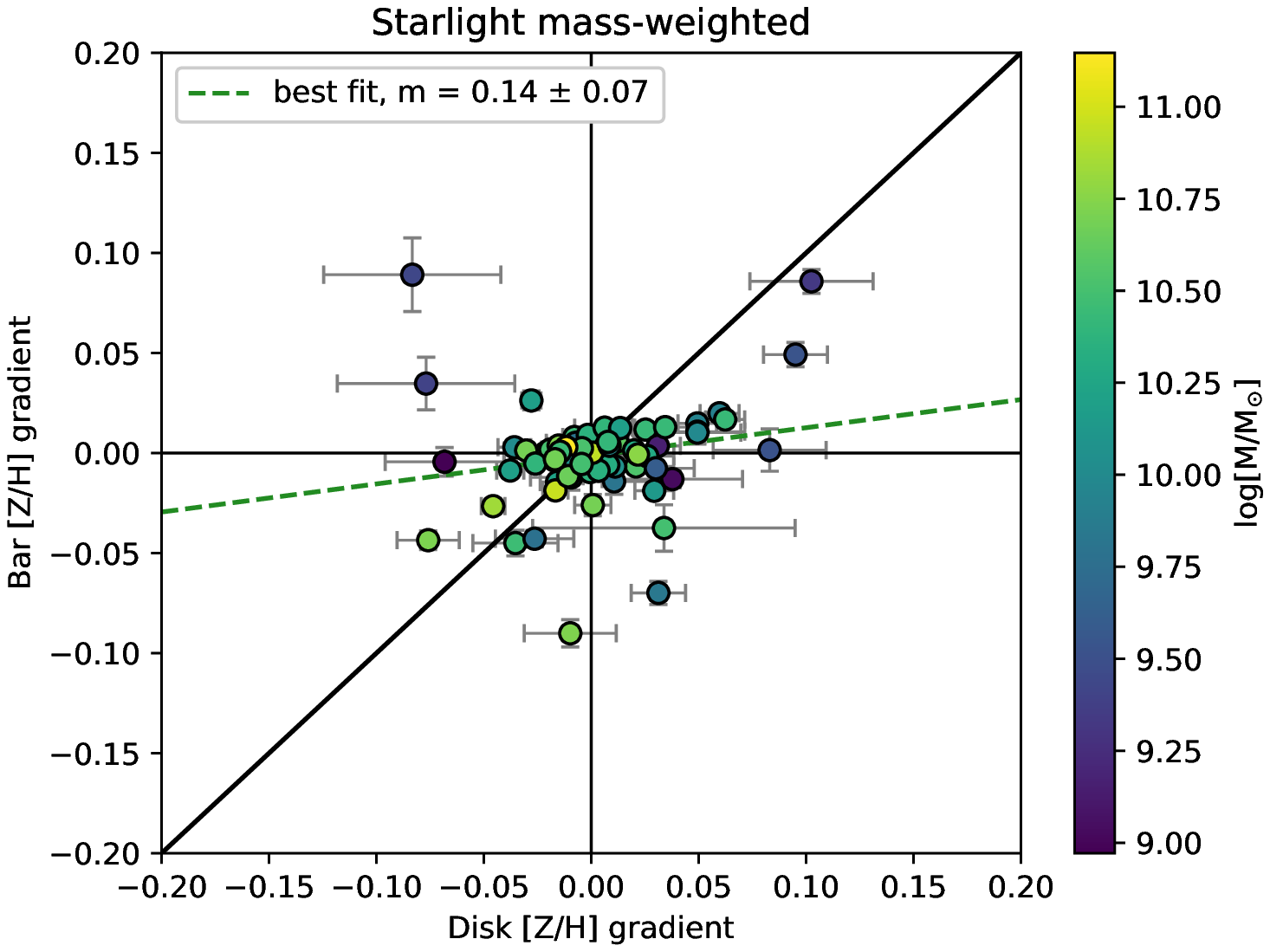} 
\end{subfigure}
\hfill
\begin{subfigure}{0.45\textwidth}
\includegraphics[width=\textwidth]{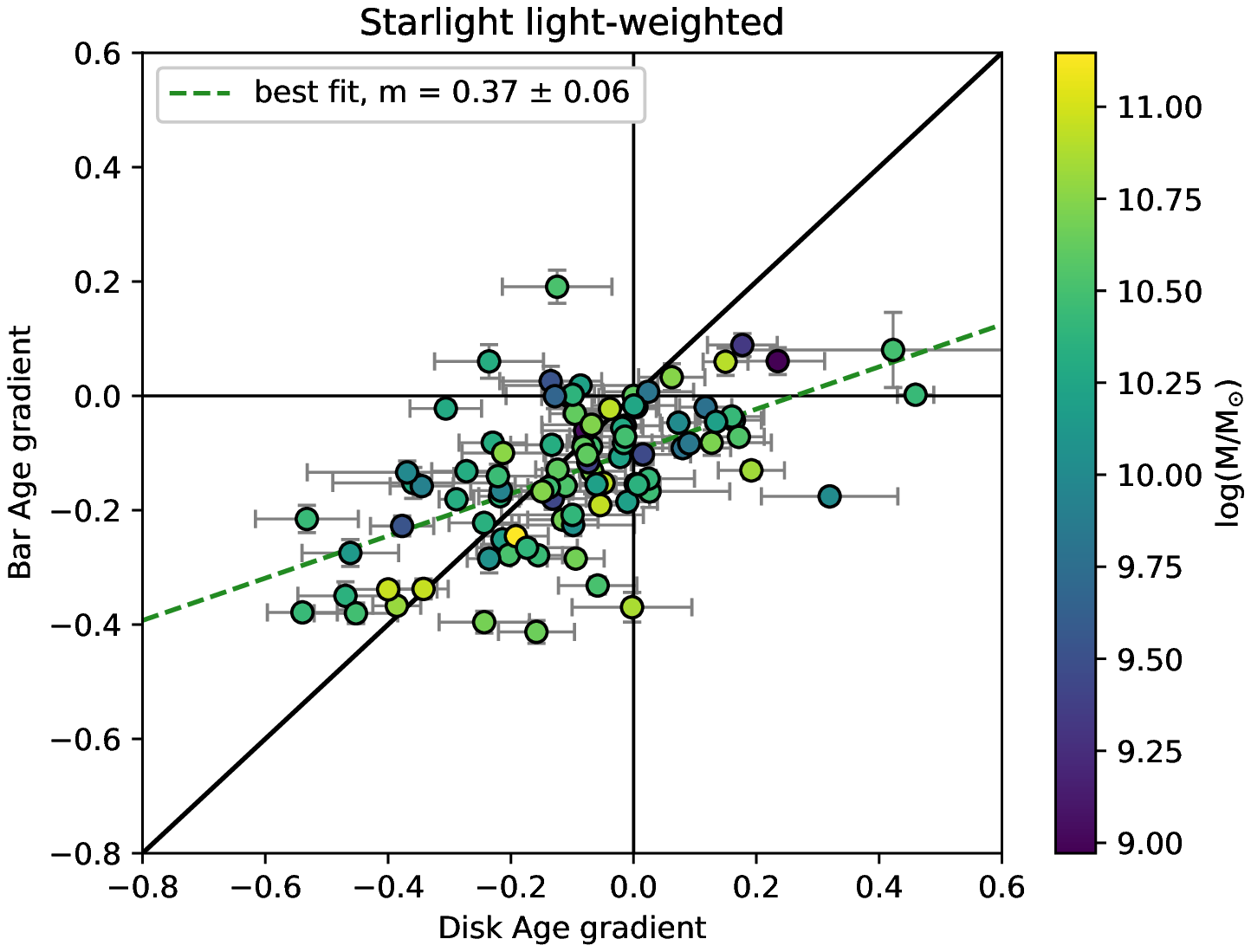} 
\end{subfigure}
\begin{subfigure}{0.45\textwidth}
\includegraphics[width=\textwidth]{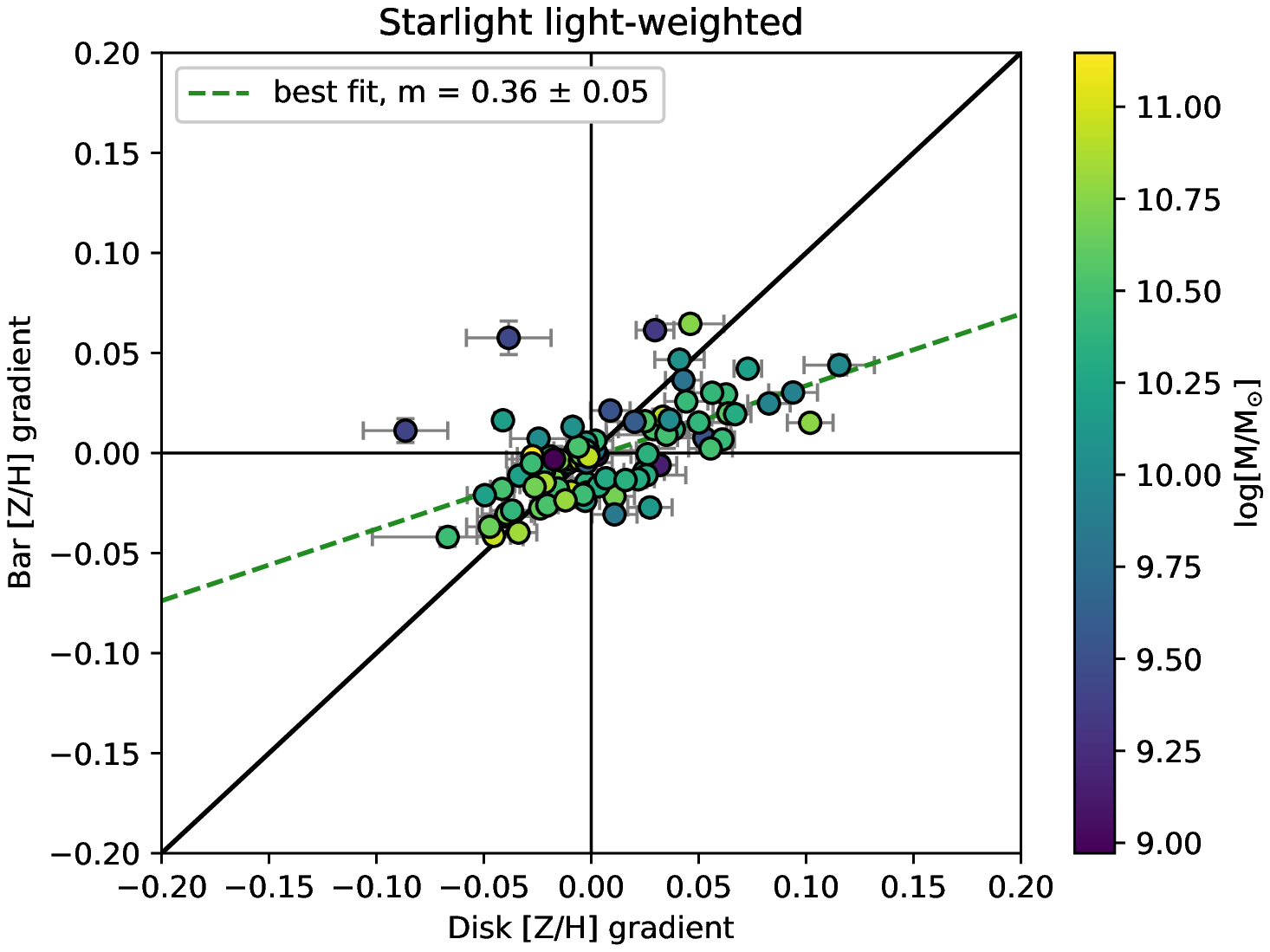} 
\end{subfigure}
\hfill
\begin{subfigure}{0.45\textwidth}
\includegraphics[width=\textwidth]{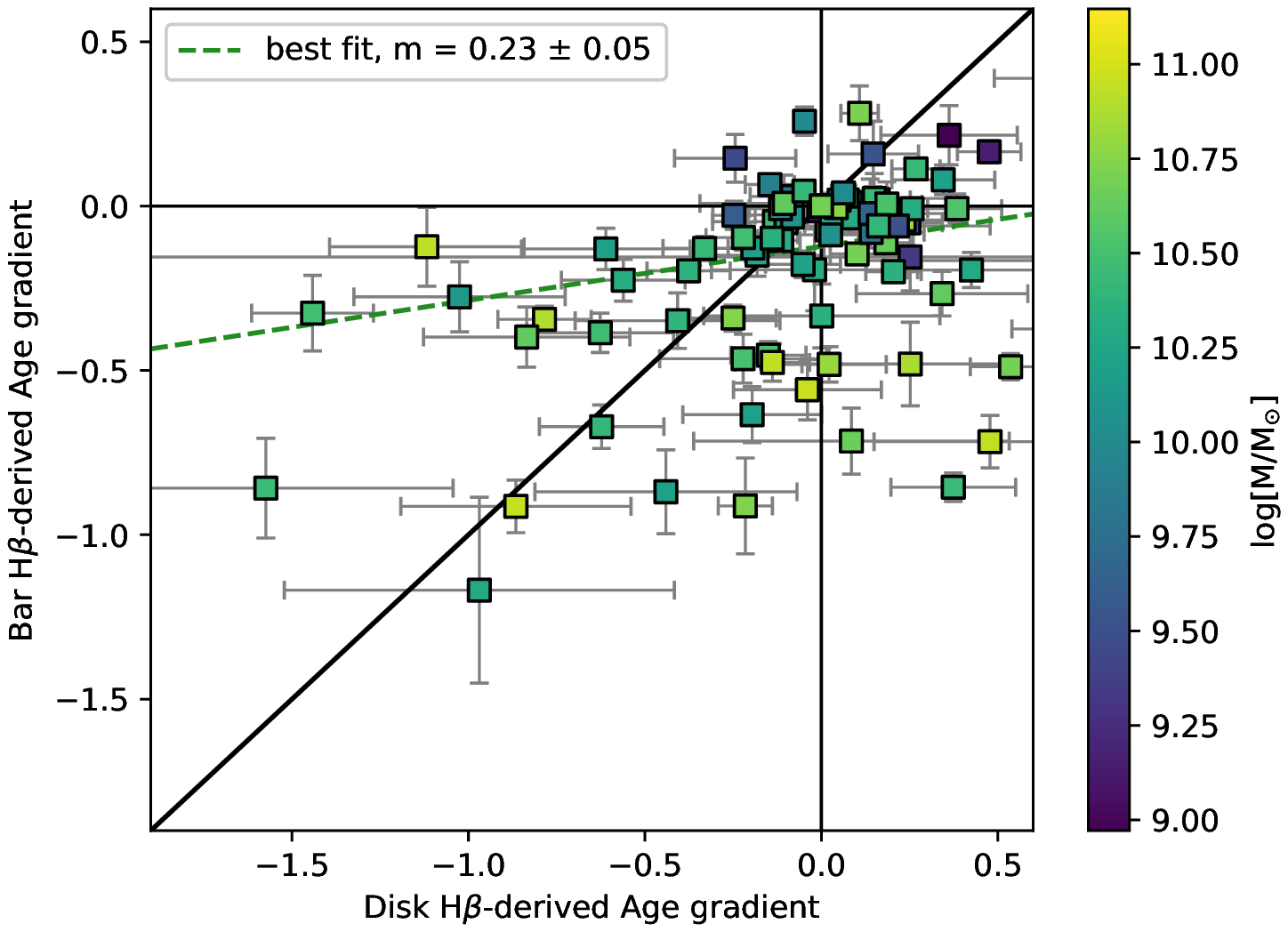} 
\end{subfigure}
\begin{subfigure}{0.45\textwidth}
\includegraphics[width=\textwidth]{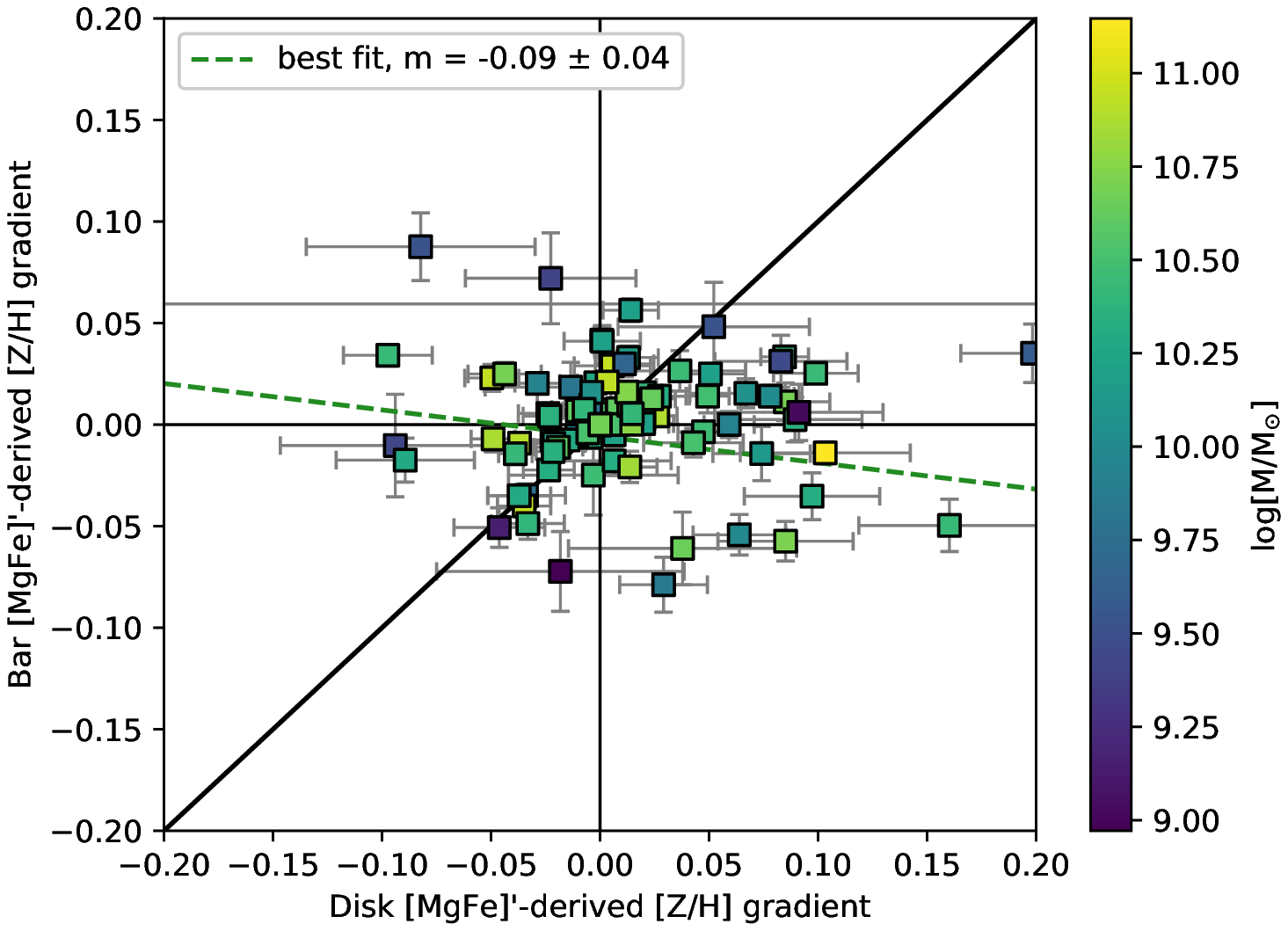} 
\end{subfigure}
\hfill

\caption{Bar and disk age (left column) and metallicity (right column) gradient comparisons for Starlight mass-weighted (top row), light-weighted (middle row) full spectral fits, and the derived ages and metallicity gradients from index measurements (bottom row, square markers, note different scale on H$\beta$-derived age plot). Black 1:1 lines denote where the bar and disk gradient is the same for a given galaxy, and a green dashed line indicates the best fit to the data points on each plot. In all cases, stellar population gradients are flatter within the bar than within the disk, indicating that bars are radially mixing stars.}
\label{gradients}
\end{figure*}

In Figure~\ref{gradients}, we present the bar and disk stellar population gradients for each galaxy in the sample, coloured by the galaxy's stellar mass, from Starlight mass-weighted and light-weighted ages and metallicities, as well as the index-derived populations (which may be thought of as being closer to the Starlight light-weighted output). 
Reassuringly, the full spectral fitting and index-derived outputs are consistent with each other. There is some scatter at the high-mass end of the H$\beta$-derived age gradients, which is likely because of the difficulty in distinguishing beyond ages of $\sim$5 Gyr using H$\beta$. 
We see that for both the age and metallicity indicators, on average, the magnitude of the gradient within the bar is significantly smaller than within the disk for all stellar masses. This is a strong indication that material is better radially mixed in the bar than within the surrounding regions. The line of best fit to the bar and disk gradients is shown in green, and the gradient, m, and 1$\sigma$ error on this value printed at the top of each panel. In each case, the best fit line slope is significantly incompatible with a gradient of unity, typically by $>5\sigma$, indicating that on average, the bar stellar population gradients are flatter than the disk measurements.
From this we conclude that bars are efficient at radially mixing material along themselves. 

While on average the bar age and metallicity gradients are flatter within the bar than the corresponding disk region, we note that some galaxies (up to 29\% of the sample for the Starlight mass-weighted age gradients) actually possess disk gradients that are flatter than the bar gradients. While this may be mostly explained by scatter caused by the error in the gradient measurements, this is not correlated with stellar mass. Further work to determine whether bar parameters (such as bar length and strength, or age), or galactic parameters (such as environment) influence bar stellar population gradients are left for a future work. We also note the portion of galaxies that possess positive metallicity gradients. Given we are excluding the very central (bulge-dominated) regions from the gradient measurement, and only measuring along the length of the bar (which is generally smaller than the radius of the galaxy), it is perhaps not surprising that gradient measurements are not all negative, and should not be compared to literature values for azimuthally-averaged galaxies over the entire radial extent of the galaxy.

It seems that the bar regions of barred galaxies possess distinct stellar population gradients compared to regions outside of the bar at the same radii. This agrees well with the results of \citet{Seidel16}, who report systematically flatter Fe5015 and Mg$b$ gradients along the bar major axis, compared to the bar minor axis for 16 barred galaxies. \citet{Seidel16} showed that the gradients found along the minor axis of the inner regions of barred galaxies are frequently comparable to the inner regions of non-barred galaxies. From this, we conclude that bars are confined structures that may be considered as an individual component within a galaxy.  
\citet{SB11} find similar when examining the long slit observations of \citet{Perez09} for two galaxies. We have shown that these trends hold robustly for a much larger sample of barred galaxies, and across a wide range of host galaxy stellar masses and bar strengths. This result is valid for bars of all evolutionary stages, which are embedded within galaxies of all stellar masses.
More generally, this analysis reinforces the importance of not azimuthally-averaging spatially-resolved data when non-axisymmetric galaxy components are present, lest subtle details be washed out.

\section{Summary \& Future Work}
We have investigated gradients in the spatially-resolved stellar populations of strongly barred galaxies in the MaNGA galaxy survey via full spectral fitting and absorption line population indicators. By separating spaxels dominated by bar light, and comparing to those within disk regions at the same radii for the first time in a large sample of galaxies, we find that:
\begin{itemize}
\item The stellar age and metallicity gradients as inferred from index measurements of H$\beta$ and [MgFe]$^{\prime}$ and Starlight full spectrum fits in the barred regions of barred galaxies are flatter than within the disk region. From this we conclude we have robust observational evidence of bars radially mixing material at all stages of bar evolution in local Universe galaxies. 

\item We confirm that individual structures within galaxies can comprise different distributions of stellar populations, and that one should not azimuthally average IFU data with non-axisymmetric structures within them as averaging loses important structural information.

\end{itemize}

Future work will involve examining how the stellar age and metallicity changes as a function of radius in barred and non-barred galaxies. We will compare this with simulations to determine how bars evolve within galaxies and the timescales involved in bar dynamics.

\section{Acknowledgements} 
The authors wish to thank the anonymous referee, whose comments improved the quality of this manuscript.
J.H.K.~acknowledges financial support from the European Union's Horizon 2020 research and innovation programme under Marie Sk{\l}odowska-Curie grant agreement No 721463 to the SUNDIAL ITN network, from the Spanish Ministry of Economy and Competitiveness (MINECO) under grant number AYA2016-76219-P, from the Fundaci\'on BBVA under its 2017 programme of assistance to scientific research groups, for the project ``Using machine-learning techniques to drag galaxies from the noise in deep imaging'', and from the Leverhulme Trust through the award of a Visiting Professorship at LJMU. 
RR thanks CNPq and FAPERGS for financial support.
Funding for the Sloan Digital Sky Survey IV has been provided by the Alfred P. Sloan Foundation, the U.S. Department of Energy Office of Science, and the Participating Institutions. SDSS-IV acknowledges
support and resources from the Center for High-Performance Computing at
the University of Utah. The SDSS web site is www.sdss.org.

SDSS-IV is managed by the Astrophysical Research Consortium for the 
Participating Institutions of the SDSS Collaboration including the 
Brazilian Participation Group, the Carnegie Institution for Science, 
Carnegie Mellon University, the Chilean Participation Group, the French Participation Group, Harvard-Smithsonian Center for Astrophysics, 
Instituto de Astrof\'isica de Canarias, The Johns Hopkins University, 
Kavli Institute for the Physics and Mathematics of the Universe (IPMU) / 
University of Tokyo, Lawrence Berkeley National Laboratory, 
Leibniz Institut f\"ur Astrophysik Potsdam (AIP),  
Max-Planck-Institut f\"ur Astronomie (MPIA Heidelberg), 
Max-Planck-Institut f\"ur Astrophysik (MPA Garching), 
Max-Planck-Institut f\"ur Extraterrestrische Physik (MPE), 
National Astronomical Observatories of China, New Mexico State University, 
New York University, University of Notre Dame, 
Observat\'ario Nacional / MCTI, The Ohio State University, 
Pennsylvania State University, Shanghai Astronomical Observatory, 
United Kingdom Participation Group,
Universidad Nacional Aut\'onoma de M\'exico, University of Arizona, 
University of Colorado Boulder, University of Oxford, University of Portsmouth, 
University of Utah, University of Virginia, University of Washington, University of Wisconsin, 
Vanderbilt University, and Yale University.
    \bibliographystyle{mnras}
  \bibliography{MaNGAbib}
\end{document}